\documentstyle[11pt,newpasp,twoside,epsf]{article} 
\def\edcomment#1{\iffalse\marginpar{\raggedright\sl#1\/}\else\relax\fi} 
\reversemarginpar 

\begin{document} 

\title{Demographics of Blazars}
\author{Giovanni Fossati} 
\affil{University of California at San Diego, 
Center for Astrophysics and Space Sciences, 
9500 Gilman Drive, La Jolla, CA 92093-0424, USA}

\begin{abstract} 
We discuss the preliminary results of an extensive effort to address the
fundamental, and yet un-answered, question that can be trivialized as:
``are there more blue or red blazars?''.
This problematic is tightly connected with the much debated issue of the
unified picture(s) of radio--loud AGNs, which in turn revolves around the
existence, and the properties of relativistic jets. 
We address this question by comparing --simultaneously-- the properties of
the collection of heterogeneously selected samples that are available now,
with the predictions of a set of plausible unifications scenarios.
We show that it is already possible to make significant progress even by
using only the present samples. 
The important role of selection effects is discussed. For instance we show
that the multiple flux selections typical of available surveys could induce
some of the correlations found in color--color diagrams. These latter
results should apply to any study of flux limited samples. 
\end{abstract} 

\section{The factor of 100 problem} 
More than 95\% of all catalogued blazars have been found in either shallow 
radio or shallow X-ray surveys (e.g. see Padovani, these proceedings). 
Because of the range of blazar spectral energy distributions (SED) the two
selection methods yield different types, the ``red'' objects 
(with the peak of the synchrotron emission at IR-optical wavelengths, LBL)
in radio samples, and the ``blue'' (whose synchrotron emission peaks at
UV-X-ray wavelengths, HBL) in X-ray samples. 
The differences in the SEDs do reflect different physical states but only
as the extrema of an underlying continuous population.

The relative space densities of the different types, not to mention their
absolute space densities or their evolution in cosmic time still remain 
indeterminate.
Different scenarios predict a difference of two orders of magnitude (!) in
the ratio of the ``red'' and ``blue'' types, nevertheless the presently
available samples are unable to distinguish between them.
The blazar demographics are this uncertain essentially because the flux
limits of current complete samples are high, so only the tip of the
population is sampled. 
The interpretation of observed phenomenology depends on the complicated
sensitivity of diverse surveys to a range of spectral types.
Ultimately, this means we do not know which kind of jets nature
preferentially makes: those with and high $B$ and $\gamma_{\rm e}$
(``blue'' blazars) or low $B$ and $\gamma_{\rm e}$ (``red'' blazars).
We also do not know whether they evolve differently and/or if ``red''
blazars dominate at high redshift and evolve into ``blue'' blazars at low
redshift, and what is the relationship between the ``non-thermal'' and
``thermal'' power/components.
The implications for understanding jet formation are obvious.

Here we present a concise account of the preliminary results of numerical
simulations of a set of unification models, including an actual fit of the
model parameters to reproduce the general characteristics of a few
reference samples (\S2).
We also introduce a ``concept'' experiment, devised to address the role of
selection effects (\S3), and discuss a couple of issues that are connected
to this problem.
In \S4 we comment on future developments.

\section{Testing unification scenarios}
We compared the existing surveys with a set of three alternative unified
schemes, following the discussion developed in recent years after Padovani
\& Giommi (1995), and Fossati et al. (1997, 1998).
They are: i) the ``radio--leading'', where the primary\footnote{%
Defined as the band where a flux limited selection would be objective with
respect to the range of intrinsic properties.} luminosity is the radio one
and N$_{\rm LBL}>$N$_{\rm HBL}$.
ii) The ``X--ray-leading'', where the primary band are the X--ray, and
N$_{\rm LBL}<$N$_{\rm HBL}$.
iii) The ``bolometric'', where the SED properties (and in turn the
distribution of L$_{\rm X}$/L$_{\rm R}$, i.e. the balance between LBL and
HBL) are determined by the total power of the source, with HBLs being the
less powerful objects.
In Fossati et al. (1997) the input parameters of each model were pre-set to
values based on those of the observed samples.  The most interesting
results was the success of the new model, the bolometric one.

\begin{figure}
\centerline{%
\epsfxsize=0.32\linewidth\epsfbox{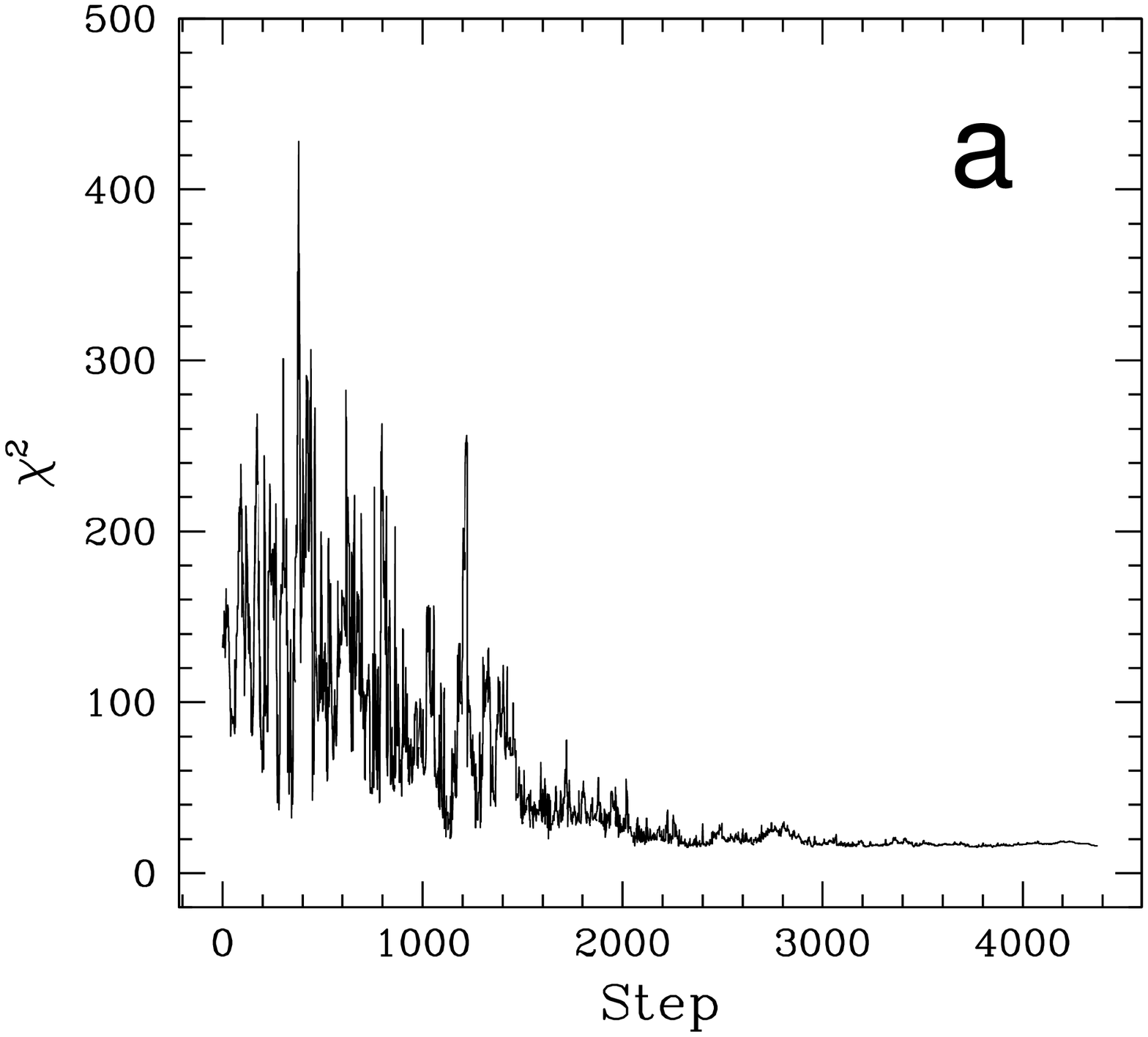}%
\epsfxsize=0.32\linewidth\epsfbox{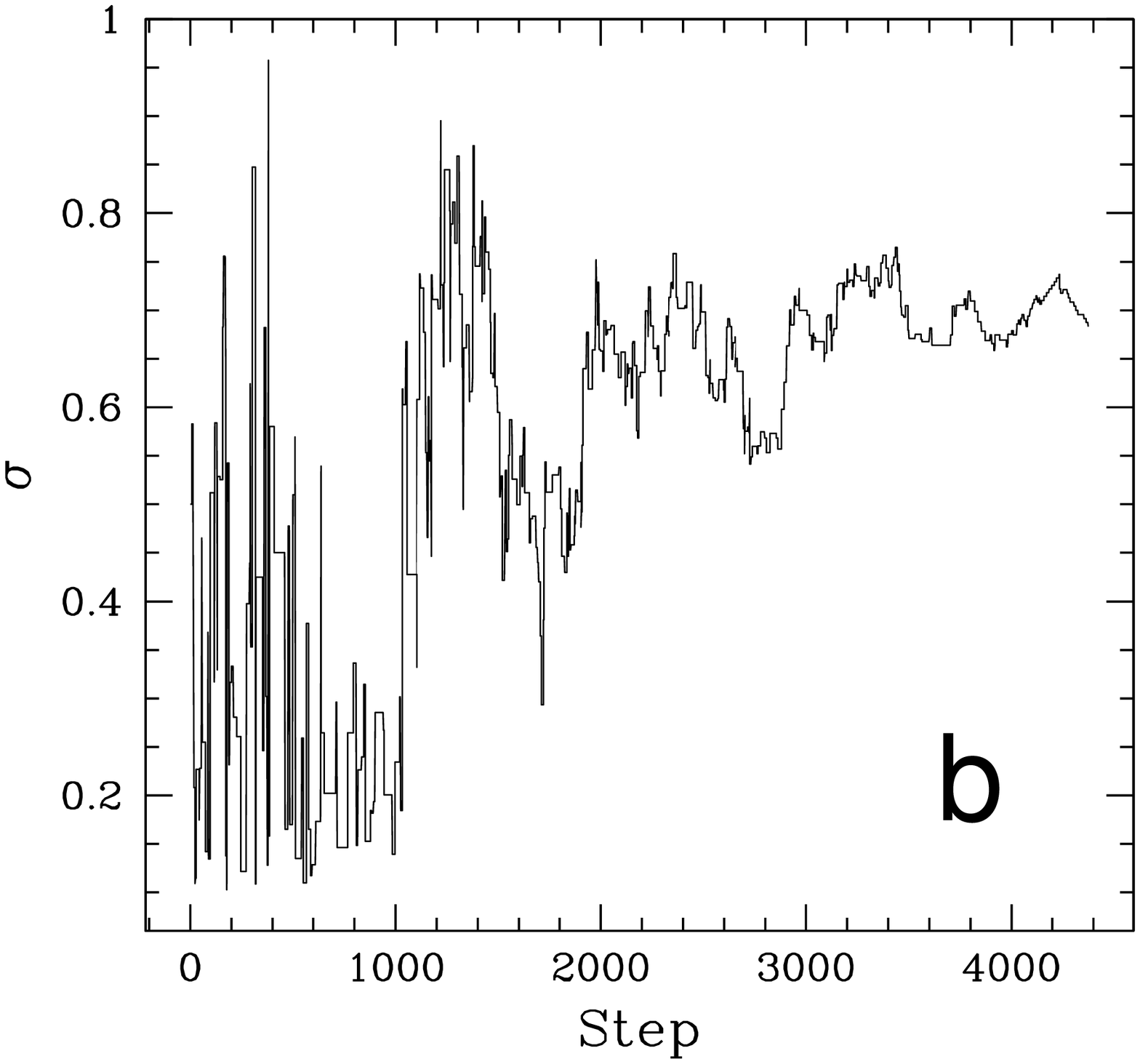}\hfill%
\epsfxsize=0.32\linewidth\epsfbox{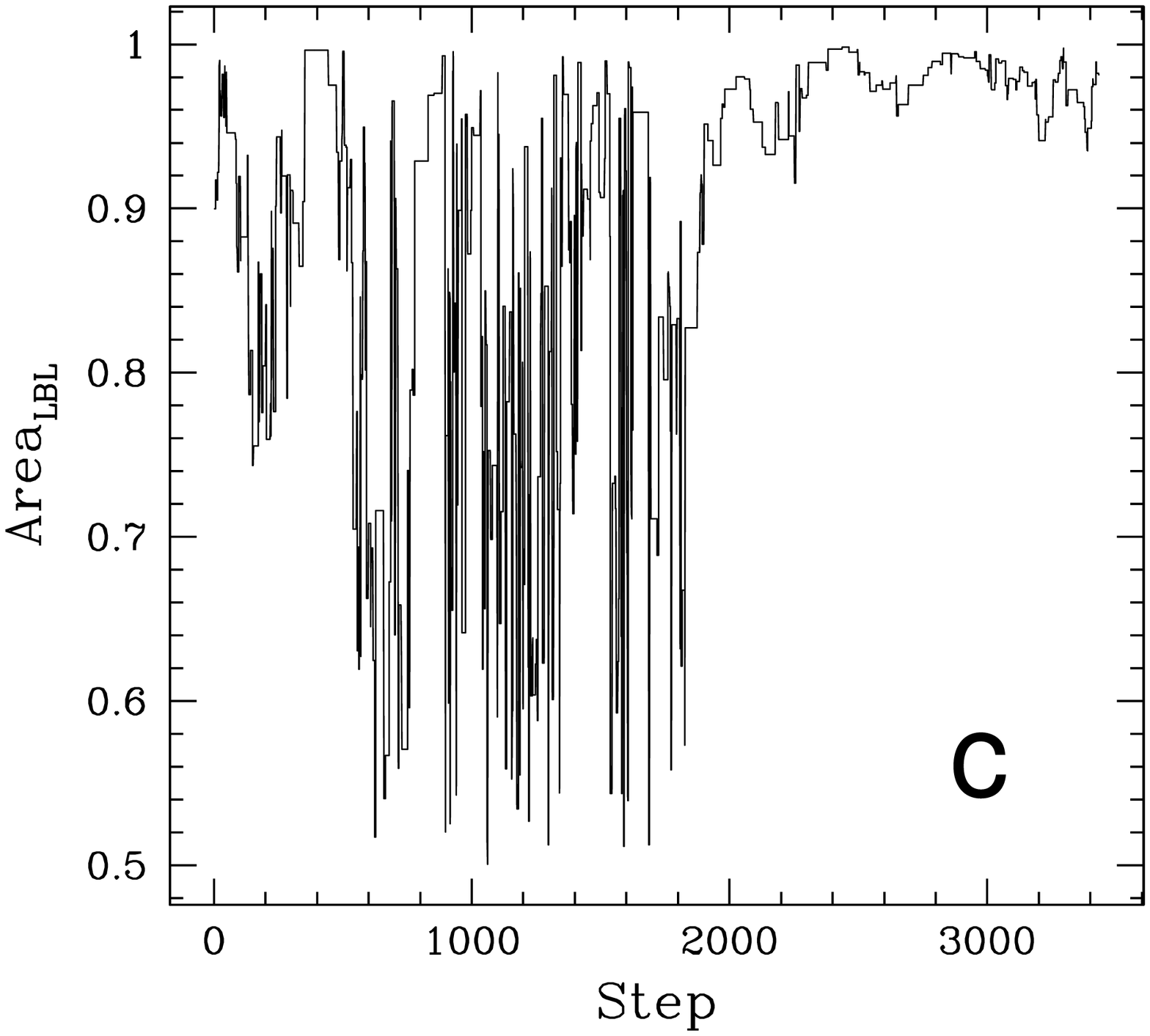}%
}
\caption{%
Evolution during the fit of the values of (a) $\chi^2$ and (b) of the width
$\sigma$ of the L--$\nu_{\rm peak}$ relationship for the bolometric model.
In (c) is plotted the area of the ``LBL Gaussian'' for the radio--leading
model.}
\end{figure}

\subsection{The fit method, and results}
In this work our approach is different.
First we normalize/optimize each unifying scheme by performing an {\it actual
fit} to three reference samples (EMSS, Slew, 1~Jy).
We leave free to vary 7--8 variables, such as the normalization and slope
of the primary luminosity function, and the distribution of the L$_{\rm
X}$/L$_{\rm R}$ ratio\footnote{%
Note on L$_{\rm X}$/L$_{\rm R}$: for the bolometric scenario we allow for a
spread in the relationship between peak frequency and luminosity.
For the radio and X--ray leading scenarios we use the combination of two
Gaussians, for which we fit the mean, sigma and area.}.
For those parameter for which there is a measured value (e.g. the
luminosity function) we allowed their values to move within their 2$\sigma$
interval.
The observational quantities to reproduce were the number, and average
radio and X--ray luminosities of HBLs and LBLs.

The technique used for the fit is {\it ``simulated annealing''}
(e.g. Kirkpatrick et al. 1983), which is based on statistical mechanics, and
implemented via MonteCarlo.
It is a very robust technique, very well suited for many parameter fits.
Moreover the ``global'' nature of the technique is very effective for cases
where there might be multiple secondary local minima in the parameter space.

In Fig.~1 we show examples of the evolution of the fit. 
We here just point out interesting results concerning two of the
``core'' issues: 
i) the best fit of the bolometric model requires a finite width for the
L--$\nu_{\rm peak}$ relationship (see Fig.~1b).
The best fit value is $\sigma \simeq 0.6$, i.e. at any given L the
synchrotron peak frequency will be distributed as a Gaussian of width
$\sigma$ centered at the $\nu_{\rm peak}$ value determined by the relationship.
ii) In both the radio-- and X--ray (not shown) leading cases the best fit
L$_{\rm X}$/L$_{\rm R}$ distribution is basically a single, broad, Gaussian
(see Fig.~1c).
For the radio--leading case the LBL Gaussian comprises 98\% of the total area,
and it is centered at $\simeq -6.3$ with $\sigma \simeq 0.6$.

\begin{figure}
\centerline{%
\epsfxsize=0.34\linewidth\epsfbox{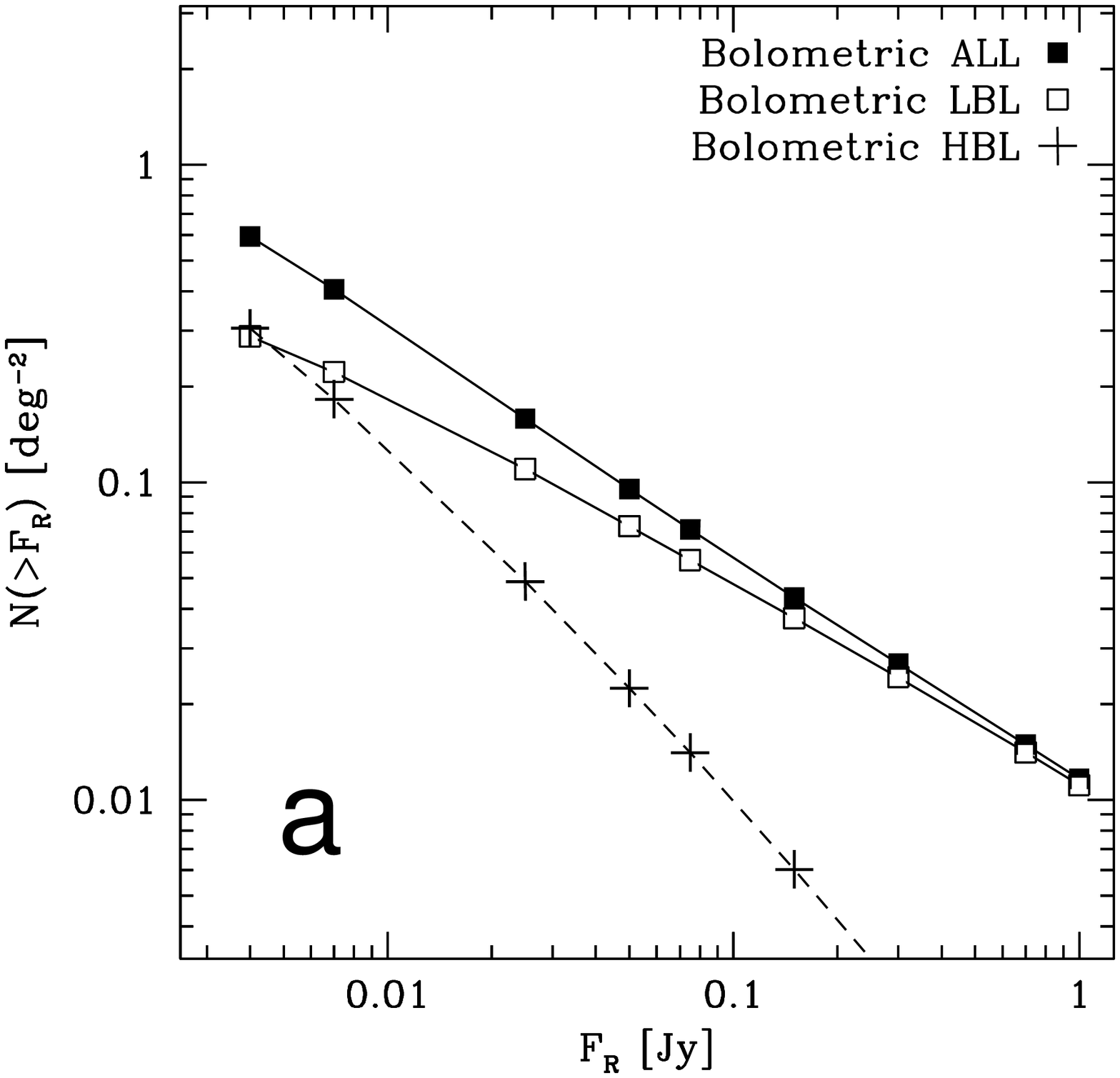}\hspace{0.02\linewidth}%
\epsfxsize=0.34\linewidth\epsfbox{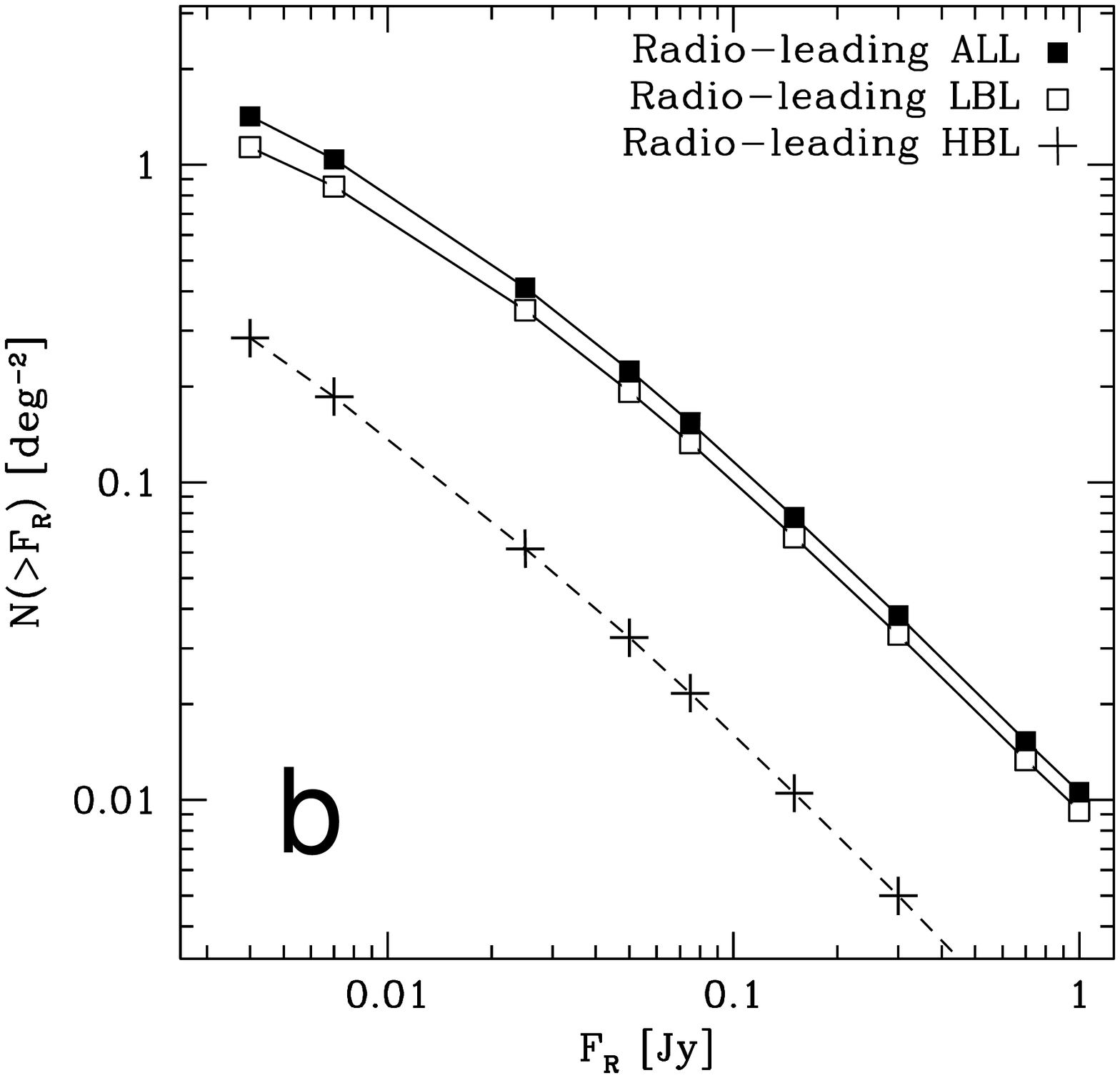}\hspace{0.02\linewidth}%
\epsfxsize=0.34\linewidth\epsfbox{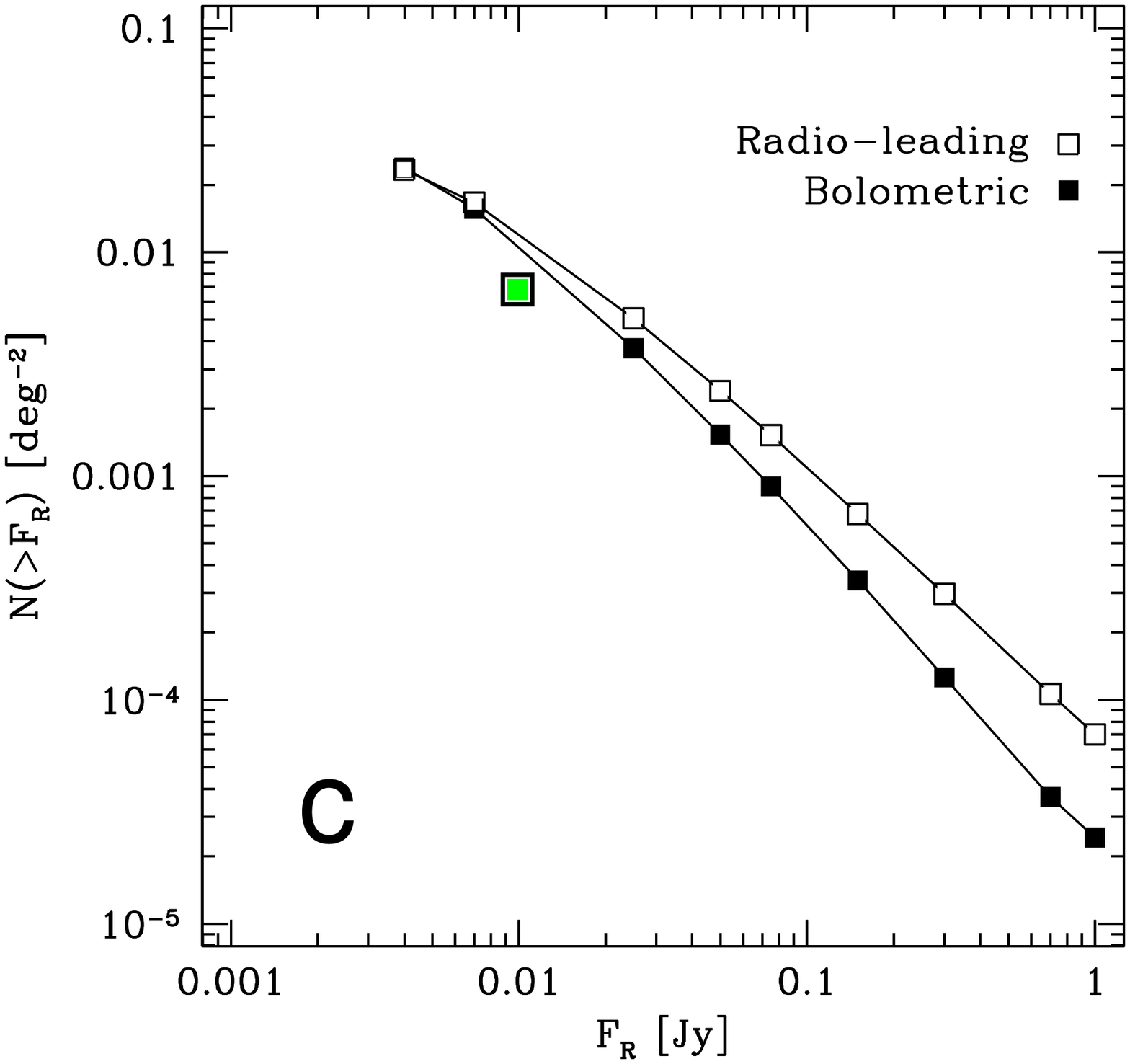}%
}
\caption{%
Radio log(N)-log(S) predicted by the (a) bolometric and the
(b) radio-leading models for the DXRBS sample.
(c) Bolometric and radio--leading predictions for the ``sedentary'' sample;
the grey-filled square represents the observed density.}
\end{figure}

\subsection{Comparison with real samples}
The next step is to use the results of the fits to predict the properties of
samples that have not been used to optimize the parameters of the models.
We present here only the integral log(N)--log(S) curves, and we only
sub-divide the samples in HBL/LBL (according to the values of F$_{\rm
X}$/F$_{\rm R}$).
It is worth noting that the absolute normalizations may not be completely
reliable, because of uncertainties on the sky coverage.  
The uncertainty on the (details of) sky coverage is indeed probably the
main one involved in the simulations.
The relative fraction of HBL and LBL may be a more robust parameter, and it
is the one more easily amenable to a quick comparison.

\subsubsection{\underbar{DXRBS}}
The DXRBS sample (Perlman et al. 1998) is still in progress, but an ``off
record'' comparison of the predicted log(N)--log(S) (shown if Fig.~2a,b)
with the observed one seems to show that the models are (still) in good
agreement with the data.
The predictions of the bolometric and radio--leading models become
radically different below about 100 mJy, a domain now reachable.
The LBL/HBL density ratio at a few radio flux limits are the following:
\begin{verbatim}
      Flux @5GHz      Bolometric  X-ray leading  Radio-leading
        @300 mJy          9.7          6.9            6.6
        @150 mJy          6.2          5.7            6.4
         @50 mJy          3.2          5.2            6.1
\end{verbatim}

\subsubsection{\underbar{Sedentary survey}}
The ``sedentary'' sample (Giommi, Menna \& Padovani 1999) comprises only
HBLs because of the built--in cut in $\alpha_{\rm RX}$. 
The radio log(N)--log(S) is shown in Fig.~2c, where the grey square
represent a density @10~mJy between the actual ``sedentary'' and the EMSS,
showing that there is a quite good agreement.
Here, as for the DXRBS, we do not plot the predictions of the X--ray
leading model because they are not satisfactory.
In fact this scenario does not seem to be able to explain the properties of
these recent samples, at least with its parameters set at the best fit values.

\begin{figure}
\centerline{%
\epsfxsize=0.43\linewidth\epsfbox{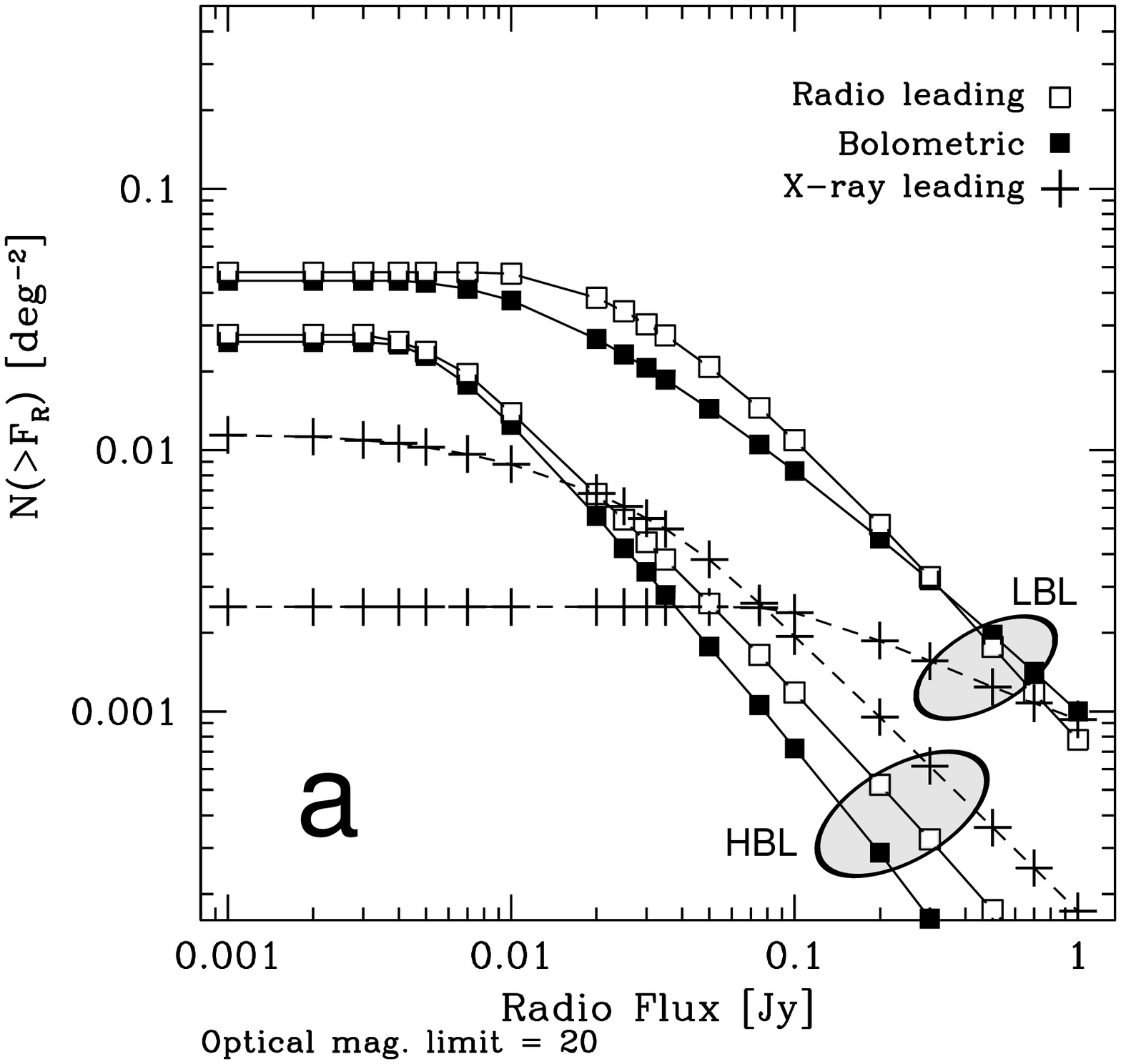}\hspace{0.08\linewidth}%
\epsfxsize=0.43\linewidth\epsfbox{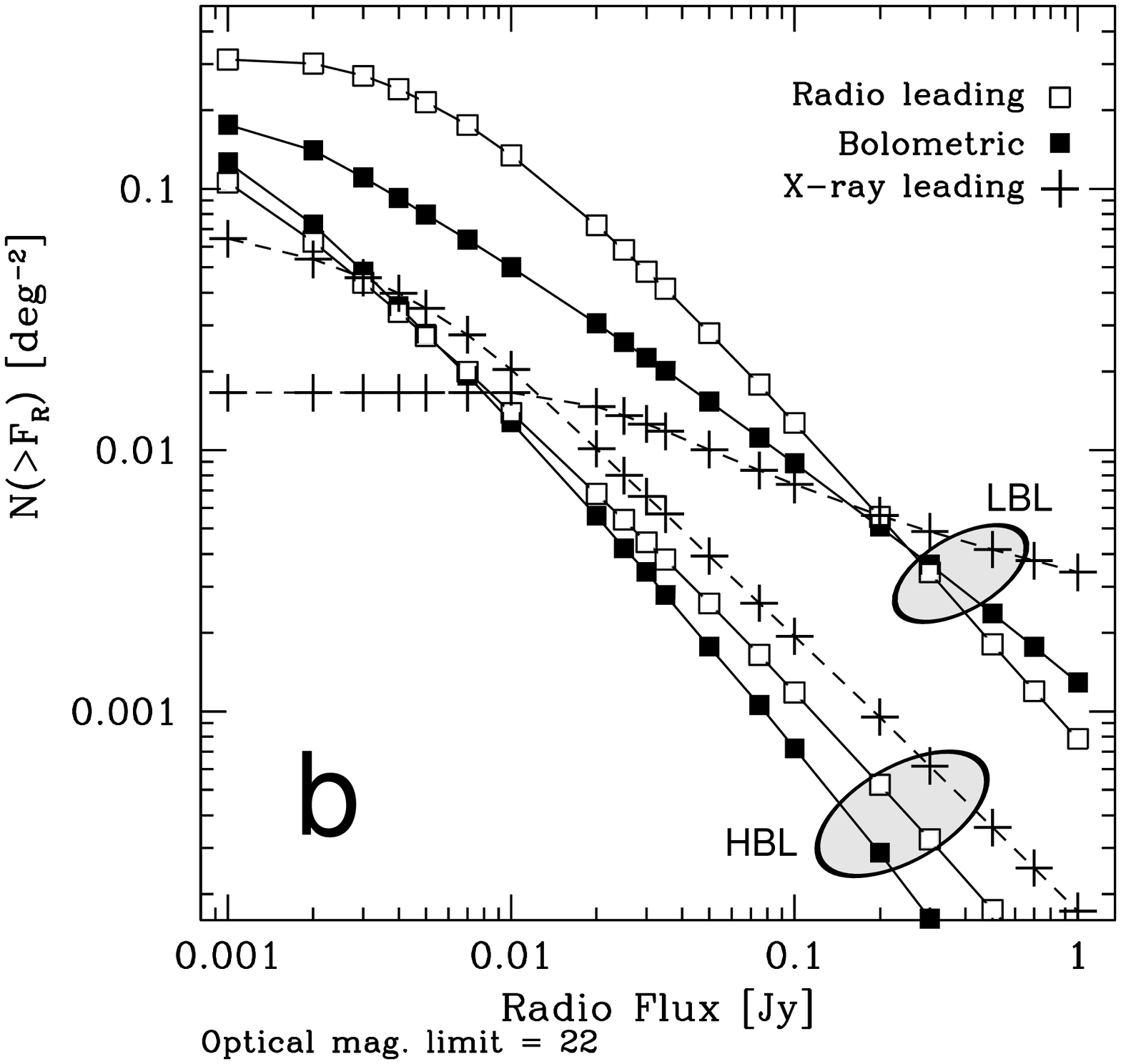}%
}
\caption{%
Radio log(N)--log(S) predicted by the bolometric/radio--/X--ray-leading
models: HBLs and LBLs for a sample with an additional cut at (a) m$_{\rm V}<$20, or 
(b) m$_{\rm V}<$22 .}
\end{figure}

\subsection{Going deeper}
In Fig.~3a,b we show the predictions of the 3 scenarios for the number
densities of HBLs and LBLs in radio surveys with a secondary cut in optical
magnitude at m$_{\rm V}$=20, and m$_{\rm V}$=22.
We see that the X--ray leading model is giving a substantially different
answer from the two other competing models, which seem to agree over most
of the accessible radio flux range.
The bolometric and radio--leading models actually start to give different
predictions only at very faint radio fluxes, as seen in Fig.~3b.
In the radio--leading model the radio counts of HBLs and LBLs keep a fixed
ratio by definition, while in the bolometric picture HBLs are deemed to
eventually outnumber the LBLs, but this seems to happen at radio fluxes
lower that expected.
However, we are not far from the range of radio fluxes that will be
the most sensitive to discriminate among the different pictures.
Actually there are already a few samples going deep enough.

\subsection{The ``cube'' (caveat \#1)}
To try to assess the problem of selection effects we introduced the ``cube''.
(Fossati \& Urry, in preparation), a toy model stripped down of every a
priori assumption as to the presumed intrinsic properties of the SED.
We assume that the radio/optical/X--ray luminosities are completely
un-correlated, and we take simple power law luminosity functions.
We then simulate samples of sources that would be selected by a generic
flux limited radio or X--ray survey (including the flux dependent sky
coverage), with a possible additional cut in another spectral band.
An example of the results of this exercise is shown in Fig~4a.
It seems to be relatively ``natural'' to obtain patterns in a color--color
diagram which looks like those that are actually observed, 
and promptly interpreted as tracing intrinsic properties of the sources.
Of course the ``cube'' is not able to reproduce the large variety of
patterns and correlations observed in luminosity--luminosity,
color--color diagrams, nevertheless we regard it as a very instructive
example of how careful we need to be when dealing with selection effects.

\begin{figure}
\centerline{%
\epsfxsize=0.40\linewidth\epsfbox{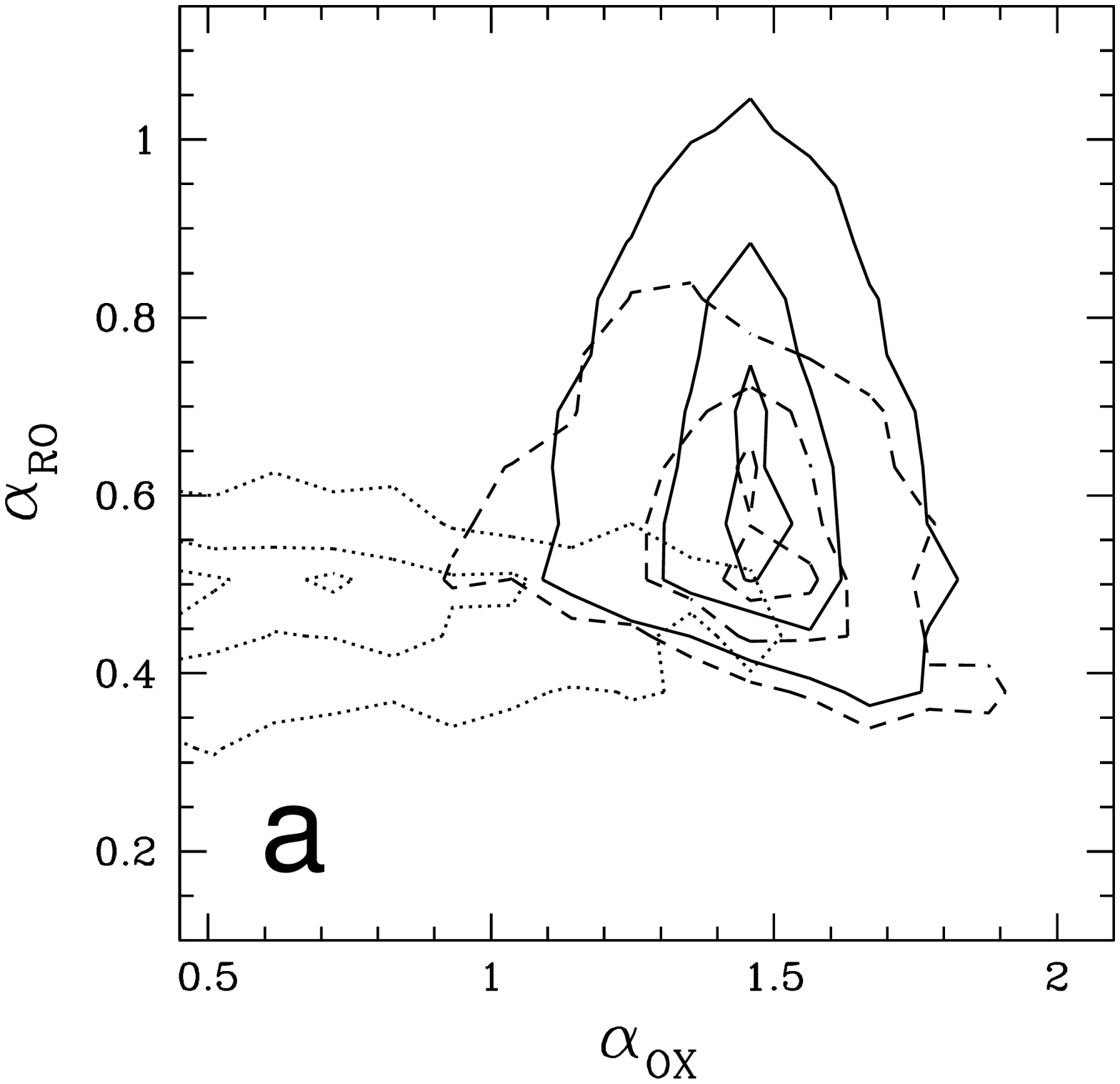}\hspace{0.10\linewidth}%
\epsfxsize=0.40\linewidth\epsfbox{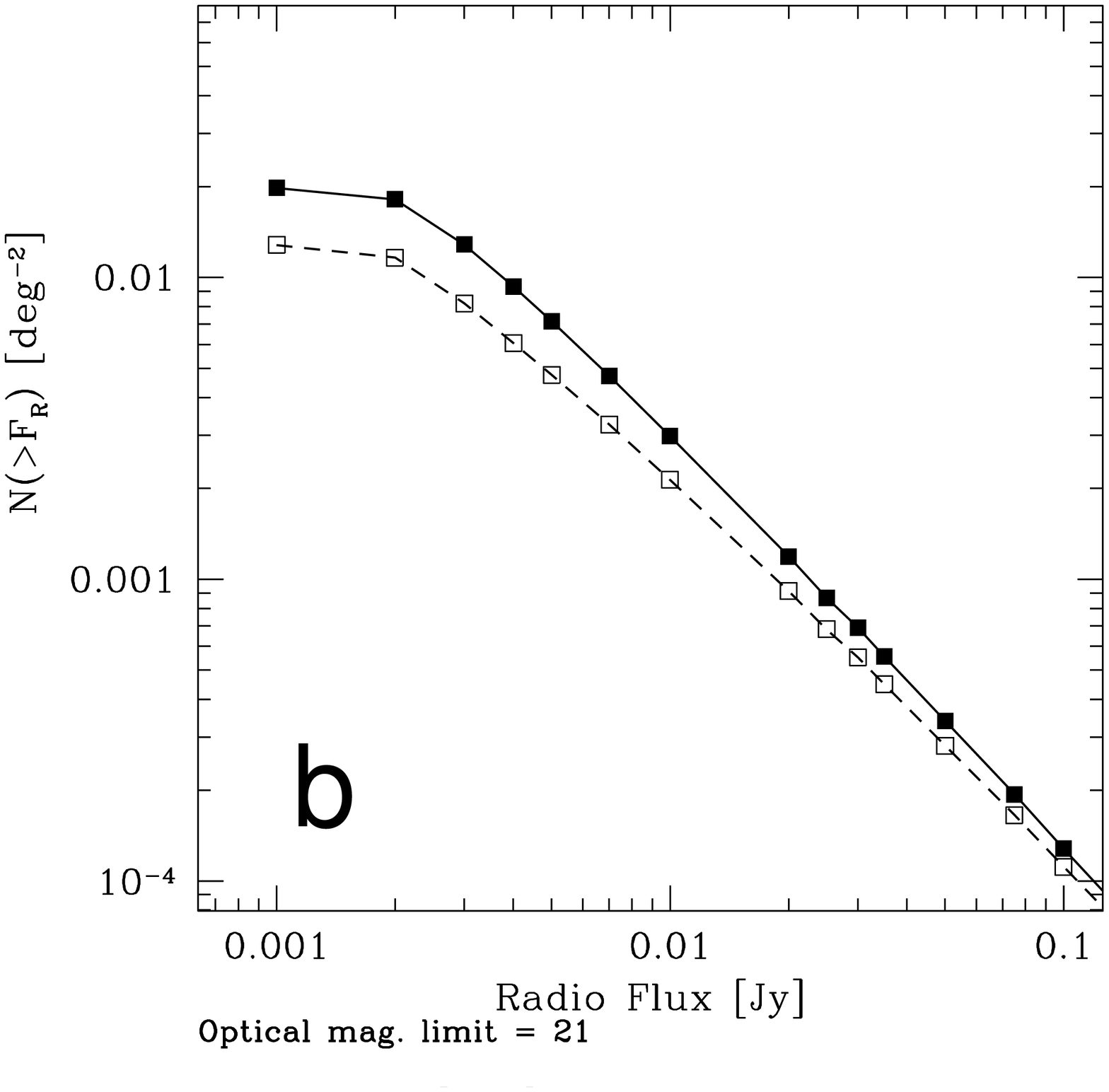}%
}
\caption{%
(a) $\alpha_{\rm RO}$ vs. $\alpha_{\rm OX}$ diagram with the density
contours predicted by the ``cube'' for an EMSS--like (dotted), a
DXRBS--like (dashed), and a radio (solid) selected samples.
(b) ``Observed'' (empty symbols) and ``intrinsic'' (filled) log(N)--log(S)
of extreme HBLs. }
\end{figure}

\subsection{Caveat \#2: on cutting in F$_{\rm X}$/F$_{\rm R}$ color}
Figure~4b shows the log(N)--log(S) of {\it observed} extreme HBL (lower
dashed line) and of {\it intrinsic} extreme HBL (upper solid line), defined
as such according to the observed or intrinsic X--ray/radio ratio.
Because of the K--correction and their SED shape, blazars systematically
shift towards the LBL side when seen at higher redshift, when
``classified'' on the basis of the {\it observed} X/radio ratio.  
The effect can be sensible when comparing relative populations of HBL and LBL.

\section{Conclusions: blazars demographics and not--so--perfect samples}
On the basis of the analysis presented here, we think that there might be
already enough information available to proceed to constrain meaningfully
the main features of unified scenarios.
The comparison of observed samples with simulations performed in a
systematic fashion (e.g. by means of simultaneous fit) may provide an
extremely powerful and effective tool to address the problem of the
intrinsic properties of blazars.

In fact, although there is not a single sample comprising all the 
desirable characteristics to provide the least possible biased picture of
the intrinsic properties of blazars, the F$_{\rm X}$/F$_{\rm R}$ plane is
now well covered (see Fig.~1, 2 in Padovani's contribution).
Moreover, the quality of the most recent samples will allow to compare the
predictions and the data directly by using the distribution of the
$\alpha_{\rm RX}$, an important step forward and past some confusion created
by selection effects combined with the ``two bins'' approach (e.g. \S2.5).

If the selection biases of each of surveys can be regarded as
being under control (and therefore reliably implemented in the simulations)
we may soon be able not only to test a given unified scheme, but even to
derive directly from the data what should be the general properties of a
successful unified scheme.

Finally, we think that more than ever it is necessary to shift the focus
away from the BL Lacs sub-class, because this could still be the source of
significant confusion.
The best progress could be made by considering the BL Lacs--FSRQs
relationship as a whole, also from the observational point of view.
The bolometric scenario was meant from the beginning to unify BL Lacs and
FSRQs, and it tries to connect some basic physical ideas to the observed
phenomenology.
On the other hand we need to figure out how to explain the HBL/LBL ratios
assumed by the radio and X--ray leading scenarios, and in turn how to extend
these models to include smoothly the FSRQs.
There should be a way to tell from ``first principles'' on which
side of the 1/10--10/1 range the real value of N$_{\rm HBL}$/N$_{\rm LBL}$
ratio is more likely to belong.  

\acknowledgements
I'd like to thank the organizers for a great workshop, and for bearing with
my request of delaying my talk by one day, and Ilaria Cagnoni for very
kindly accepting to swap our talks in the schedule.  I also thank {\it
pippol} for the neverending support.


\begin{references}
\reference Fossati, G. et al. 1997, \mnras, 289, 136
\reference Fossati, G. et al. 1998, \mnras, 299, 433
\reference Giommi, P., Menna, M.T., \& Padovani, P. 1999, \mnras, 310, 465
\reference Kirkpatrick, S. et al. 1983, Science, 220, 671
\reference Padovani, P., \& Giommi, P. 1995, \apj, 444, 567
\reference Perlman, E., et al. 1998, \aj, 115, 1253
\end{references}
\end{document}